\documentclass[letter]{aa} 

\usepackage{graphicx}
\usepackage{txfonts}
\usepackage{orcidlink}
\usepackage{bm}
\usepackage[font=small, skip=0pt]{caption}
\usepackage{natbib,twoopt}
\usepackage[hyphenbreaks]{breakurl}
\usepackage{float}
\DeclareMathAlphabet\mathzapf       {T1}{pzc} {mb} {it}
\bibpunct{(}{)}{;}{a}{}{,}             
\definecolor{cobalt}{rgb}{0.06, 0.2, 0.65}
\hypersetup{
  colorlinks,
  citecolor=cobalt,
  linkcolor=cobalt,
  urlcolor=cobalt,
}
\makeatletter
  \newcommandtwoopt{\citeads}[3][][]{\href{http://adsabs.harvard.edu/abs/#3}%
    {\def\hyper@linkstart##1##2{}%
     \let\hyper@linkend\@empty\citealp[#1][#2]{#3}}}
  \newcommandtwoopt{\citepads}[3][][]{\href{http://adsabs.harvard.edu/abs/#3}%
    {\def\hyper@linkstart##1##2{}%
     \let\hyper@linkend\@empty\citep[#1][#2]{#3}}}
  \newcommandtwoopt{\citetads}[3][][]{\href{http://adsabs.harvard.edu/abs/#3}%
    {\def\hyper@linkstart##1##2{}%
     \let\hyper@linkend\@empty\citet[#1][#2]{#3}}}
  \newcommandtwoopt{\citeyearads}[3][][]%
    {\href{http://adsabs.harvard.edu/abs/#3}
    {\def\hyper@linkstart##1##2{}%
     \let\hyper@linkend\@empty\citeyear[#1][#2]{#3}}}
\makeatother

\newcommand{\swift}{{\em Swift}}

\newcommand{\ixpe}{{\em IXPE}}

\newcommand{\be}{\begin{equation}}
\newcommand{\en}{\end{equation}}

\def\ltsima{$\; \buildrel < \over \sim \;$}
\def\lsim{\lower.5ex\hbox{\ltsima}}
\def\gtsima{$\; \buildrel $\geq$ \over \sim \;$}
\def\gsim{\lower.5ex\hbox{\gtsima}}

\def\arc{\mbox{$^{\prime\prime}$}}

\def\deg{\mbox{$^{\circ}$}}

\def\lum {\mbox{erg~s$^{-1}$}}
\renewcommand{\linenumbers}{}
\begin{document} 


   \title{Multi-band study of the flaring mode emission in the transitional millisecond pulsar PSR\,J1023+0038}

   \author{M.~C. Baglio 
          \inst{1}
          \and
          F.~Coti Zelati\inst{2, 3, 1} 
          \and
          A.~K.~Hughes \inst{4,5} 
          \and
          F.~Carotenuto \inst{5,6} 
          \and
          S.~Campana \inst{1} 
          \and
          D.~de~Martino\inst{7}
          \and       
          S.~E.~Motta \inst{1}
          \and
          A.~Papitto \inst{6}   
          \and
          N.~Rea\inst{2,3}
          \and
          D.~M.~Russell \inst{8}
          \and
          D.~F.~Torres\inst{2,3,9}
          \and
          A.~Di~Marco\inst{10}
          \and    
          F.~La~Monaca\inst{10}
          \and
          S.~Covino\inst{1, 11}
          \and
          S.~Giarratana \inst{1}
          \and
          G.~Illiano\inst{6}
          \and
          A.~Miraval~Zanon\inst{12}
          \and
          K.~Alabarta\inst{8}
          \and
          P.~D'Avanzo \inst{1}
          \and
          M.~M.~Messa\inst{13,1}
          }

   \institute{INAF--Osservatorio Astronomico di Brera, Via Bianchi 46, I-23807 Merate (LC), Italy\\
              \email{cristina.baglio@inaf.it}
         \and
             Institute of Space Sciences (ICE, CSIC), Campus UAB, Carrer de Can Magrans s/n, E-08193 Barcelona, Spain
          \and
          Institut d'Estudis Espacials de Catalunya (IEEC), 08860 Castelldefels (Barcelona), Spain
          \and 
          Department of Physics, University of Alberta, CCIS 4-181, Edmonton, AB T6G 2E1, Canada
          \and
          Astrophysics, Department of Physics, University of Oxford, Keble Road, Oxford OX1 3RH, UK
          \and
          INAF-Osservatorio Astronomico di Roma, Via Frascati 33, I-00076, Monte Porzio Catone (RM), Italy
          \and
    INAF--Osservatorio Astronomico di Capodimonte, Salita Moiariello 16, I-80131 Naples, Italy
          \and
          Center for Astrophysics and Space Science, New York University Abu Dhabi, PO Box 129188, Abu Dhabi, UAE
          \and
    Institució Catalana de Recerca i Estudis Avançats (ICREA), Passeig Lluís Companys 23, E-08010 Barcelona, Spain
    \and
    INAF Istituto di Astrofisica e Planetologia Spaziali, Via del Fosso del Cavaliere 100, I-00133 Rome, Italy
    \and
Como Lake centre for AstroPhysics (CLAP), DiSAT, Universit\'{a} dell’Insubria, via Valleggio 11, 22100 Como, Italy
    \and
    ASI - Agenzia Spaziale Italiana, Via del Politecnico snc, I-00133 Rome, Italy
\and
    Dipartimento di Fisica, Università degli Studi di Milano, Via Celoria 16, I-20133 Milan, Italy
    }

   \date{Received XX; accepted XX}

 
 \abstract{
We present a comprehensive study of the flaring mode of the transitional millisecond pulsar (tMSP) PSR J1023+0038 during its X-ray sub-luminous state, using strictly simultaneous X-ray, UV, optical, and radio observations. The X-ray flares exhibit UV and optical counterparts and coincide with the brightest radio flare observed in the past decade, reaching 1.2\,mJy at 6\,GHz and lasting $\sim$1\,hour. During the flare, the optical polarisation drops from $\simeq$1.4\% to $\simeq$0.5\%, indicating the emergence of an unpolarised component. We propose that the thickening of the disc, which enlarges the shock region between the pulsar wind and the accretion flow and may drive the X-ray flaring observed in tMSPs, enhances the ionisation level of the disc, thereby generating an increased number of free electrons. These electrons could then be channelled by magnetic field lines into the jet. This increased jet mass-loading could drive the associated radio and optical variability. The radio spectral evolution during flares is consistent with synchrotron self-absorption in jet ejecta or internal shocks within the compact jet. We infer radio polarisation upper limits ($<$8.7\%, $<$2.3\%, and $<$8.2\%, before,
during, and after the radio flare) that further support a compact jet origin but do not rule out discrete ejections. Our findings suggest that tMSPs could serve as essential laboratories for investigating jet-launching mechanisms, mainly because they operate under very low mass accretion rates. This accretion regime has not been explored before in the context of accretion-ejection coupling.
}


   \maketitle
%

\section{Introduction}
\label{sec:intro}

PSR J1023+0038 (J1023) is the archetypal transitional millisecond pulsar (tMSP). It was classified as a tMSP following the detection of radio pulsations revealing a 1.69-ms spin period neutron star (NS) in a 4.75-hour binary system \citep{Archibald2009}. In 2013, J1023 underwent a sudden multi-wavelength flux increase, accompanied by the disappearance of its radio pulsations and the appearance of optical emission lines. This change marked a transition to a peculiar state characterised by the presence of an accretion disc and an average X-ray luminosity of 10$^{33-34}$\,\lum, known as the X-ray sub-luminous state \citep{Stappers2014}. In this state, J1023 shows three X-ray luminosity modes \citep{Linares2014ApJ,Archibald15,Bogdanov15,cotizelati18,papitto19}: high ($\sim$7$\times$10$^{33}$\,\lum, 70--80\% of the time); low ($\sim$10$^{33}$\,\lum, 20--30\%); and flaring ($\gtrsim$10$^{34}$\,\lum, 1--2\%). X-ray, UV, and optical pulsations at the NS spin period are detected only in the high mode and in flaring mode, even though at low significance (\citealt{papitto19,miravalzanon22}; see also \citealt{Ambrosino17,jaodand21}). Radio and millimetre observations reveal bright continuum emission with an anti-correlation between the radio and X-ray bands: higher radio flux is observed during the low mode, with the spectrum shifting from flat ($\alpha \sim 0$, where $\alpha$ is the spectral index) to inverted \citep[$\alpha>0$;][]{Bogdanov2018}. The radio emission is attributed to synchrotron radiation originating from a \cite{blandford79}-type `compact jet' in the high mode and discrete ejecta in the low mode \citep[][see also \citealt{Baglio2023}]{fender04}. 

To explain the behaviour of J1023, recent scenarios propose the presence of an always-active radio pulsar \citep{papitto19,Veledina19,Baglio2023}. In the high mode, pulsed X-ray/UV/optical emission originates from synchrotron-emitting regions in a shock whereby the pulsar wind interacts with the inner accretion flow, $\sim$100-200\,km from the pulsar. 
In the low mode, the shock may move outwards, reducing the X-ray flux and causing the disappearance of optical-to-X-ray pulsations \citep{papitto19}. Alternatively, during low modes, the in-falling matter may penetrate within the light cylinder, pushing J1023 into a propeller regime with lower dissipation, reducing the X-ray luminosity and suppressing pulsations \citep{Veledina19}. \citet{Baglio2023} detected rapid millimetre flares during the high-to-low mode switch; they attribute them to the ejection of the inner accretion flow as optically thin ejecta, leading to the observed increase in the radio flux during the low mode. 

In the high mode, synchrotron emission from the shock contribute $\simeq$5\% of the optical flux and $\simeq$90\% of the soft X-ray flux \citep{Baglio2023}. The shock region emits polarised radiation at levels of $\approx$16\% in X-rays (in the high mode) and $\approx$1\% in the optical band (on average), with aligned polarisation angles. For low- and flaring-mode data, X-ray polarisation upper limits of $26$\% and $28$\%, respectively, were derived. Radio polarisation was not detected, with upper limits of $\sim$10\% in the high mode and $\sim$19\% in the low mode \citep{Baglio2024}.

Despite extensive studies on low and high modes, the flaring mode is less explored, and most research has focused solely on its X-ray evolution. Current scenarios \citep{papitto19, Veledina19} suggest that the flaring mode is due to an increase in the thickness of the inner accretion disc intercepting the pulsar wind, though clear observational evidence is lacking. This Letter presents the first multi-band study of the flaring mode of J1023 aimed at providing a comprehensive understanding of the behaviour of tMSPs in this emission mode in the frame of the scenarios described above.

\section{Data analysis}

Table\,\ref{tab:log} reports the observations presented in this work. 
Data were barycentered at the Solar System using the DE440 ephemeris and the position: R.A.=10$^\mathrm{h}$23$^\mathrm{m}$47$^\mathrm{s}$69; Dec.=+00$^{\circ}$38$^{\prime}$40.$^{\prime\prime}$8 (J2000.0; \citealt{deller12}).
\subsection{IXPE}
\label{sec:ixpe}
The Imaging X-ray Polarimetry Explorer (\ixpe) observed J1023 in two sessions from May 29 to June 14, 2024, totalling $\simeq$675\,ks of exposure per detector unit (DU). An in-depth analysis of these data was reported by \cite{Baglio2024}. After reducing instrumental background in the data \citep{DiMarco2023}, we selected source photons from a 60\arc\ radius circle, and background photons from an annulus with radii of 150\arc\ and 300\arc. We extracted the combined background-subtracted light curve from the three DUs using \texttt{ixpeobssim} \citep{Baldini2022} and \texttt{FTOOLS}.

\subsection{Swift}
\label{sec:swift}
We performed two observations with the \emph{Neil Gehrels Swift} Observatory (\swift), employing the X-ray Telescope (XRT; \citealt{burrows05}) in photon-counting mode and the Ultra-Violet/Optical Telescope (UVOT; \citealt{roming05}) in event mode with the UVM2 filter. 
We screened XRT data using standard procedures and collected source photons from a circle with a radius of 47.2\arc\ 
and background photons from an annulus with radii of 94.4\arc\ and 188.8\arc centred on the target. We extracted background-subtracted time series in the 0.5--10\,keV range, with 50-s bins. 
For the UVOT data, we used a circle with a 5\arc\ radius to collect source photons, and a circle away from the source with a 10\arc\ radius for background photons. We extracted background-subtracted time series binned at 30\,s using the \texttt{uvotevtlc} tool.

\subsection{VLT}
\label{sec:vlt}
\begin{figure*}
  \sidecaption
  \includegraphics[width=0.68\textwidth, trim=0 10 0 10, clip]{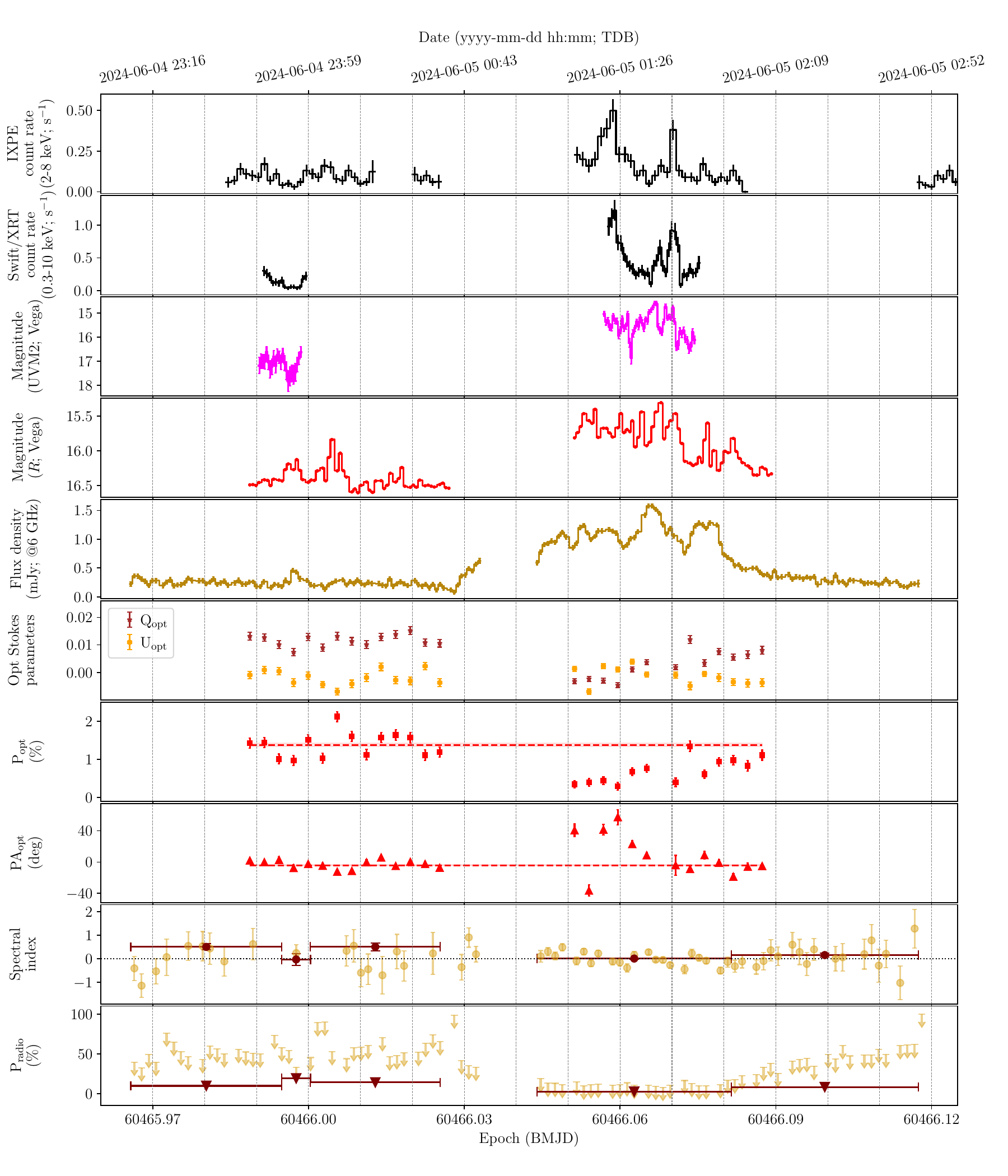}
  \captionsetup{font=small, skip=0pt}
  \caption{Multi-band time series and evolution of optical and radio parameters (including data presented by \citealt{Baglio2024}). Error bars represent the 1$\sigma$ confidence levels. In the seventh and eighth panels, horizontal dashed red lines mark the average optical polarisation degree (P$_{\rm opt}$) and angle (PA$_{\rm opt}$), with shaded areas indicating one standard deviation about the mean. In the ninth panel, a horizontal dotted line shows a spectral index of 0 (i.e., a flat spectrum). In the final panel, downward arrows denote 3$\sigma$ upper limits on the radio polarisation degree (P$_{\rm radio}$). TDB stands for barycentric dynamical time.}
  \label{fig:lcurves}
\end{figure*}

J1023 was observed with FORS2 on the Very Large Telescope (VLT) at Cerro Paranal, Chile, on June 4--5, 2024, for 7742\,s ($\simeq$2.15 hours), covering 45\% of its orbital period. The observations, split into two sets (3946\,s and 3796\,s), produced 112 images with 20\,s integration each.
We obtained normalised Stokes parameters, $Q_{opt}$ and $U_{opt}$, for linear polarisation and we evaluated the intrinsic polarisation level, $P_{\rm opt}$, and angle, $\theta_{\rm opt}$, following \cite{Baglio2020, Baglio2023}. Details are given in Appendix\,\ref{pol_analysis}. The photo-polarimetric data also allowed us to extract the system $R$-band light curve, since the sum of $f^{o}(\Phi)$ and $f^e(\Phi)$ for each image gives the total target flux. We did this calculation for J1023 and two field stars to check for any anomaly and used the light curves of the field stars to perform differential photometry. The resulting light curve of J1023 is shown in Fig.\,\ref{fig:lcurves}.

\subsection{VLA}
\label{sec:vla}
Two radio observations of J1023 were conducted in the C band ($\sim$4--8\,GHz) with the \emph{Karl G. Jansky} Very Large Array (VLA). The first observation was conducted on June 4, 2024, simultaneously with \ixpe, \emph{Swift}, and the VLT. 
The second observation was conducted on September 15, 2024. The observations were calibrated with the Common Astronomy Software Applications VLA pipeline (\texttt{casa} v6.5, \citealt{Casa2022}). The June observations included polarisation calibrators; as a result, following the pipeline, we manually solved for polarisation calibration solutions. We discuss the details of the observations, imaging, and data analysis in Appendix \ref{sec:Appendix_vla}.

\section{Results}
\label{sec:results}
\subsection{Multiband photometry}

J1023 entered the X-ray flaring mode during an observation gap between barycentric modified Julian day (BMJD) 60466.025 and 60466.052. The first \ixpe\ data point post-gap already has a count rate above the typical values in the low and high modes. After an initial flare lasting $\approx$15 min, the X-ray emission drops before exhibiting more flares. J1023 then resumes alternating between low and high modes after around BMJD 60466.07483.

During the first VLA observation, J1023 underwent a bright ($\gtrsim$\,1mJy), rapidly evolving flare that lasted for $\sim$1\,hr; before and after the flare, the source was significantly dimmer ($\sim$\,240\,$\mu$Jy; see Figs.\,\ref{fig:lcurves} and \ref{fig:spindex_flare}).
Notably, the radio flux begins rising around BMJD 60466.03, peaks around BMJD 60466.06, and declines to pre-flare levels by BMJD 60466.09. The radio flare is complex, with substructures suggesting repeated events, and the spectral index shifts from optically thick to optically thin on minute timescales. Qualitatively, the radio flares appear to exhibit a small frequency delay, with the evolution at 5\ GHz lagging behind that at 7\ GHz (see Fig.~B.1, middle panel). To quantify the delay, we calculated the cross-correlation function between the sub-bands using the \texttt{ZDCF} algorithm \citep{zdcf}, measuring the uncertainty on the most significant lag with \texttt{PLIKE} \citep{plike}. The ZDCF analysis favoured a non-zero lag of $0.7_{-0.4}^{+0.3}$\,min, consistent with our interpretation of a propagating synchrotron emitting plasma (i.e. a jet), though the significance of the lag is marginal (i.e. $\sim$1.8$\sigma$; Fig.\,\ref{fig:flare_delay}). 

Alongside the radio flaring, the optical and UV fluxes increase significantly. The first optical data point after the gap is $\sim$0.6 magnitudes brighter than before. Similarly to the radio band, the optical flare may peak around BMJD 60466.06, but its shorter duration complicates a comparison with the radio flare. The optical light curve shows two small flares early on, peaking at BMJD 60465.997 and BMJD 60466.005, lasting $\sim$7 and $\sim$10 minutes, respectively. The first flare has a radio counterpart (peaking at BMJD 60465.997 and lasting $\sim$10\,min) and corresponds to an X-ray low mode (Fig. \ref{fig:lcurves}).


\subsection{Multi-band polarimetry}
Figure \ref{fig:lcurves} shows the time evolution of the optical and radio polarisation properties of J1023. The average optical polarisation fraction is $(1.38\pm0.03)\%$ before the bright flare, decreasing to $(0.49\pm0.04)\%$ between BMJD 60466.051 (the first flare observation) and BMJD 60466.071, before returning to its pre-flare level during the decay. Pre-flare, the Stokes parameters, $Q_{\rm opt}$ and $U_{\rm opt}$, are approximately constant. $U_{\rm opt}$ remains constant during the flare, but $Q_{\rm opt}$ evolves, initially decreasing before returning to its pre-flare value by BMJD 60466.071. The average polarisation angle is $PA_{\rm opt}$ = $-1.0\deg\pm 0.9\deg$ before the flare. Similar to the polarisation fraction, the polarisation angle changes during flaring and returns to its pre-flare value during the flare decay.

No significant radio polarisation was detected in our observations. Imaging on 1-min intervals allowed us to place $3\sigma$ upper limits of ${\lesssim}\,55\%$ on average on the polarisation fraction before and after the radio flare, and of ${\lesssim}\,10\%$ during the flare. Imaging on longer time intervals improved the upper limits to ${<}\,8.7\%$, ${<}\,2.3\%$, and ${<}\,8.2\%$, before, during, and after the radio flare. We observed short-timescale variability of the spectral indices, suggesting that the significance of synchrotron self-absorption varies during the sub-flaring. This is reminiscent of the `optically thick flares' observed in X-ray binaries \cite[XRBs; e.g.][]{Fender2019}, where the rise is optically thick and the decay is optically thin. Given that optically thin synchrotron radiation is more polarised than its optically thick counterpart, we also imaged by combining the times when the radio flux density was rising and decaying (green and yellow regions; Fig.~\ref{fig:spindex_flare}). In both images, J1023 remains undetected with upper limits of ${<}\,3.7\%$ and ${<}\,3.6\%$ during rise and decay times.

\section{Discussion}\label{sec:Origin of the flaring behaviour}

The observed X-ray flaring has UV, optical, and radio counterparts. Previous studies reported two types of X-ray flares: short (seconds to minutes), isolated flares, and flares happening over extended periods (hours), with intense optical and UV flares coinciding with both types of X-ray flare (\citealt{Bogdanov15}; see also \citealt{kennedy18,papitto19,miravalzanon22}). Flares as long as 10 hours have also been detected  \citep{Tendulkar2014}. \cite{jaodand21} observed quasi-periodic flares in X-ray, UV, and optical bands, each with an average duration of $\approx$500\,s. This suggests a mechanism that is distinct from high-low mode switching and that continues to operate during flares alongside additional broadband emission. Conversely, radio flares exhibit a more complex connection with X-ray activity, occurring during low-mode episodes, accompanying X-ray flares, or appearing independently \citep{Bogdanov2018}.

In the June VLA dataset, we detected the brightest ($\gtrsim$1.5\,mJy at 6\,GHz) and longest ($\sim$50 minutes) radio flaring from J1023 since its 2013 state transition. This prominent radio flare is likely unrelated to high-low mode switches, being much longer and brighter than those typically seen during low modes. For comparison, a smaller radio flare around BMJD 60465.997 during an X-ray low mode lasted $\sim$5 minutes and reached $\sim$0.46 mJy. The radio flare displays clear substructures, with multiple sub-flares superimposed on an increased baseline radio flux density. A similar substructure is observed in the optical light curve.

The optical polarisation degree decreases during the flare, and the polarisation angle rotates before gradually reverting to its pre-flare value as the flaring activity subsides (Figs.\,\ref{fig:lcurves}, \ref{fig:flux_pol}). This indicates that a polarised component dominates when the flare component is absent or faded, and that an unpolarised component emerges during the flare. Following \cite{Baglio2024} and noting the alignment of optical and X-ray polarisation angles, we propose that the polarised component originates from the shock between the pulsar wind and the inner accretion flow, which is then dominated by the unpolarised flaring emission.

\bigskip
The origin of the observed multi-band flaring behaviour is not well understood. However, the simultaneity of the X-ray, UV, optical, and radio flaring suggests a common underlying mechanism. In tMSPs, X-ray flares likely result from transient increases in inner accretion disc thickness caused by thermal instabilities or magnetic field rearrangements \citep{Veledina19}. A thicker disk could intercept a larger fraction of the pulsar wind, enhancing energy dissipation and irradiation and thereby increasing the X-ray luminosity. The enhanced irradiation may boost the population of free particles at the light cylinder radius. If these particles reach the jet, the resulting increase in mass-loading would produce a more luminous jet and, hence, stronger optical and radio emission. Strong short-timescale optical variability is a typical feature of jets in XRBs (e.g. BH XRBs GX\,339-4, V404 Cygni, MAXI J1535-571, and MAXI J1820+070; \citealt{Gandhi2010, Gandhi2017, Baglio2018, Paice2019, Tetarenko2021}). In this scenario, the optical flaring from the weakly polarised compact jet dominates over the shock radiation, leading to depolarisation. The jet likely contributes little to the flaring UV emission, which may arise from the same mechanism as X-ray flares (i.e. thickening of the shock region), increasing dissipation and fluxes. Similarly, optical flares might also originate in the disc rather than the jet. In this interpretation, the decreased optical polarisation may result from less ordered magnetic fields in the shock due to inner disc thickening, while the change in polarisation angle could reflect a change in the projection of the average magnetic field direction of the shock on the sky plane.

The radio variability holds an insight into the nature of the jet(s). Similar to what is observed in J1023, compact jets in XRBs are known to flare on minute-to-hour timescales, resulting in light curves that are delayed at lower frequencies \citep[e.g.][]{tetarenko21}. In one interpretation, the radio variability is due to shocks between discrete shells of material moving at different velocities along the jet axis \citep[due to stochastic fluctuations in the mass-loading from the accretion flow, e.g.][]{malzac14}. The delays result from propagation times, whereby different frequencies probe distinct regions along the jet axis (higher frequencies are closer to the jet base).

Alternatively, the spectral variability and frequency delays during the radio flaring resemble the optically thick flares typically seen in XRBs and are commonly associated with the launching of discrete jet ejecta \cite[e.g.][]{Fender2019}. In this scenario, the radio spectrum evolves from self-absorbed to optically thin during the flare rise and decay. Indeed, the behaviour of the radio spectral indices is broadly consistent with this trend, as the maximum ($\alpha\,{\sim}\,0.5$) spectral index occurred as the flux density rose, and the minimum ($\alpha\,{\sim}\,-0.5$) occurred when the flux density decayed. This is supported by our integrated images, in which we combined times when the source was rising and decaying to boost the signal-to-noise ratio (yellow and green regions on Fig.~\ref{fig:spindex_flare}). We observe a statistically significant difference in the spectral indices during the rise ($\alpha=0.18\,{\pm}\,0.04$) and decay ($\alpha=-0.16\,{\pm}\,0.04$). Smaller flares from J1023 during low-mode episodes have also been interpreted as discrete ejections \citep{Bogdanov2018, Baglio2023}. 

We searched for evidence of jet ejecta in our September follow-up VLA observations $\sim$\,3 months after the initial multi-wavelength campaign. While we did not detect any ejecta, with a $3\sigma$ upper limit on the flux density of $\sim$7.5\,$\mu$Jy, this does not rule out that the observed flaring is connected to jet ejections, as the ejecta may be short-lived, similar to those seen in the BH XRB V404 Cygni \citep{tetarenko17,millerjones19}. Moreover, J1023 was, once again, exhibiting radio flaring (see Fig.\,\ref{fig:VLA_SEP}), reaching a maximum flux density of $\sim$1.2\,mJy. As a result, slow-moving ejecta (i.e. $\lesssim\,20$mas/day) would blend into the bright, flaring core emission, making them undetectable. The absence of polarised radio emission ($\lesssim$2$\%$) also seems to favour a compact jet origin for the flaring. XRB compact jets regularly exhibit polarisations $\lesssim$\,3$\%$ \citep{Han92,Corbel2000,Russel2015}, while ejecta can approach the theoretical limit for optically thin synchrotron radiation \citep[$\sim$50\%;][]{Brocksopp2013,curran14}. However, some XRBs launch ejecta with polarisation fractions well below that threshold, such as the 
jets from V404 Cygni \citep{Han92,Hughes23A}. 

By associating each optically thick-to-thin spectral transition from J1023 with a single ejection, we can estimate the minimum energy of each ejecta using standard minimum energy calculations under equipartition conditions and assuming that the
flare spectrum evolution is driven by decreasing optical depth to synchrotron self-absorption \citep{Fender2019}. We identify at least five individual flares with such transitions (Fig.\,\ref{fig:spindex_flare}). We estimate minimum energies of $\simeq (1.4-2.2) \times 10^{35}$\,erg, corresponding to a minimum total kinetic feedback of $\approx$10$^{36}$\,erg within 2 hours, and an equipartition magnetic field of $\simeq$1.4\,G per ejection. Assuming each ejection has a duration consistent with the flare rising timescale (e.g. \citealt{Carotenuto_2024}), $\simeq$300\,s for J1023, we estimate a jet power of $\simeq$$(5 - 8) \times 10^{32}$\,\lum\ ($\sim$1--10\% of the X-ray luminosity during flares; \citealt{Bogdanov15}). 


The radio properties of J1023 closely resemble those of V404 Cygni during its 2015 outburst, including short-timescale (<hour) radio flaring \citep{tetarenko17}, rapidly evolving spectral indices \citep{fender23}, and very low polarisation levels (0.05–0.3\% at 6 GHz; \citep{Hughes23A}). 
VLBI confirmed that V404 Cygni’s flaring was linked to short-lived jet ejecta \citep{millerjones19}, with multi-wavelength modelling revealing lower jet power than typical XRBs \citep[${\sim}\,(4 - 3000)\times 10^{32}$\,\lum;][]{tetarenko17}. J1023’s jet power overlaps with the lower range of V404 Cygni’s ejecta. VLBI observations of J1023’s flaring could confirm jet ejecta, with follow-up studies investigating how J1023 and other tMSPs power ejections despite differing geometry and lower accretion power (see Appendix \ref{sec:radiative_lum}).

\section{Conclusions}
\label{sec:conclusions}
Our observations of J1023 have enabled the most comprehensive study of the flaring mode of a tMSP in a sub-luminous state to date. The X-ray flaring has UV and optical counterparts and coincides with the brightest radio flare observed for J1023 after the 2013 state transition. During the flare, the optical polarisation drops, indicating the emergence of an unpolarised component. While X-ray and UV flares may result from the thickening of the inner accretion disc, we argue that the jet could drive the optical and radio flaring. The lack of correlation between the UV and optical light curves supports this interpretation. The larger dissipation may generate more free electrons, contributing to loading the compact jet. The radio properties resemble compact jet flaring seen in XRBs, though the variability and spectral evolution suggest that discrete ejecta may also contribute. The evolution of the radio-band spectral index during flares is consistent with synchrotron self-absorption in jet ejections or shocks in the compact jet. The derived radio polarisation upper limits ($< 8.7\%$, $< 2.3\%$, and $< 8.2\%$, before, during, and after the radio flare) favour a compact jet origin but do not rule out discrete ejecta.

The detection of J1023 in a bright radio state during two brief observations conducted months apart suggests that the source may have entered a previously unobserved phase characterised by frequent radio flaring. Continued monitoring at radio wavelengths could provide further insights into this behaviour. VLBI observations during subsequent flaring may conclusively identify whether the flares result from the launching of discrete ejecta. If confirmed, tMSPs may become crucial laboratories for probing jet launching mechanisms in distinct accretion regimes. 

\bigskip
\noindent
\textbf{Data Availability:} The reduced FORS2/VLT data are available at DOI: \href{https://doi.org/10.5281/zenodo.14840659}{10.5281/zenodo.14840659}. Raw data can be obtained from ESO archive upon request (\url{http://archive.eso.org/}). This research used data products provided by the \ixpe\ Team (MSFC, SSDC, INAF, and INFN) and distributed with additional software tools by the High-Energy Astrophysics Science Archive Research Center (HEASARC), at NASA Goddard Space Flight Center (GSFC).
The data that support the findings of this study are publicly available at their respective online archive repositories (\swift: \url{https://heasarc.gsfc.nasa.gov/cgi-bin/W3Browse/w3browse.pl};  
VLA: \url{https://data.nrao.edu/portal/}).

\begin{acknowledgements}
MCB is supported by the INAF-Astrofit fellowship.
FCZ is supported by a Ram\'on y Cajal fellowship (RYC2021-030888-I). 
AKH is supported by NSERC Discovery Grant RGPIN-2021-0400. 
FC is supported by the Royal Society through the Newton International Fellowship programme (NIF/R1/211296).
SC and PD'A are supported by ASI grant I/004/11/5.
ADM and FLM are supported by the Italian Space Agency (ASI) via contract ASI-INAF-2022-19-HH.0.
AP and DdM are supported by ASI and INAF under agreements ASI-INAF I/037/12/0, ASI-INAF n.2017-14-H.0, INAF Sostegno alla ricerca, and INAF SKA/CTA projects. AP acknowledges support from the Fondazione Cariplo/Cassa Depositi e Prestiti, grant no. 2023-2560.
FCZ, SCa, PD'A, AP, DdM, and GI are supported by the INAF-FANS project and MUR PRIN 2020 grant GEMS 2020BRP57Z.
DMR and KA are supported by Tamkeen under the NYU Abu Dhabi Research Institute grant CASS.
DFT is supported by the grant PID2021-124581OB-I00 funded by MCIU/AEI/10.13039/501100011033 and 2021SGR00426.
GI is supported by the AASS Ph.D. program with INAF collaboration.
NR is funded by ERC Horizon 2020 grants MAGNESIA (No. 817661) and DeepSpacePulse (No. 101189496).
This work is supported by the Spanish Unidad de Excelencia Mar\'ia de Maeztu CEX2020-001058-M and MCIU with EU NextGeneration EU funding (PRTR-C17.I1).
Based on observations collected under ESO programme 113.27RE.001.
\end{acknowledgements}

\bibliographystyle{aa}
\bibliography{biblio}

\newpage
\begin{appendix}

\section{Journal of the observations}
Table\,\ref{tab:log} reports details on the observations presented in this work.

\begin{table*}
\caption{Observations of J1023.}
\label{tab:log}
\centering
\resizebox{2.0\columnwidth}{!}{
\begin{tabular}{lcccccc}
\hline\hline
Telescope/Instr.& Obs./Prg. Id    & Setup         & Start -- End time                             & Exposure  & Band      & Ref.\\
                        &                 &               & YYYY Mmm DD hh:mm:ss (UTC)                     & (ks)      &           & \\
\hline
\ixpe\           & 03005599       &                  & 2024 May 29 12:30:51  -- 2024 Jun 14 14:57:45  & 675 & 2--8\,keV    & \cite{Baglio2024}  \\
\swift/XRT/UVOT  & 00033012239    & PC / Event mode  & 2024 Jun 04 22:14:59 -- 2024 Jun 04 23:59:54  & 1.5  & 0.3-10\,keV / UVM2  & \cite{Baglio2024}  \\
\swift/XRT/UVOT  & 00033012240    & PC / Event mode  & 2024 Jun 05 01:23:09 -- 2024 Jun 05 01:48:53  & 1.5  & 0.3-10\,keV / UVM2   & \cite{Baglio2024} \\
VLT/FORS2       & 113.27RE.001    & Wollaston prism + & 2024 Jun 04 23:34:37 -- 2024 Jun 05 00:39:23 & 3.9    &  $R$ & \cite{Baglio2024} \\
                &                 & HWP              & 2024 Jun 05 01:06:34 -- 2024 Jun 05 02:08:50   & 3.8    &  $R$  & This work \\
VLA             & 24A-476         & B configuration  & 2024 Jun 04 23:10:30  -- 2024 Jun 05 02:49:21   & 11.1    & $C$ & \cite{Baglio2024} \\
                &                 & B configuration  & 2024 Jun 04 23:10:28  -- 2024 Jun 05 02:49:22   & 13.2    & $C$  & This work \\
VLA             & 24B-469         & B configuration  & 2024 Sep 15 14:51:18 -- 2024 Sep 15 17:29:27  & 7.65  & C   & This work \\
\hline
\end{tabular}
}
\end{table*}

\section{Polarimetry with VLT/FORS2}\label{pol_analysis}
\begin{figure}
\begin{center}
\includegraphics[width=0.5\textwidth]{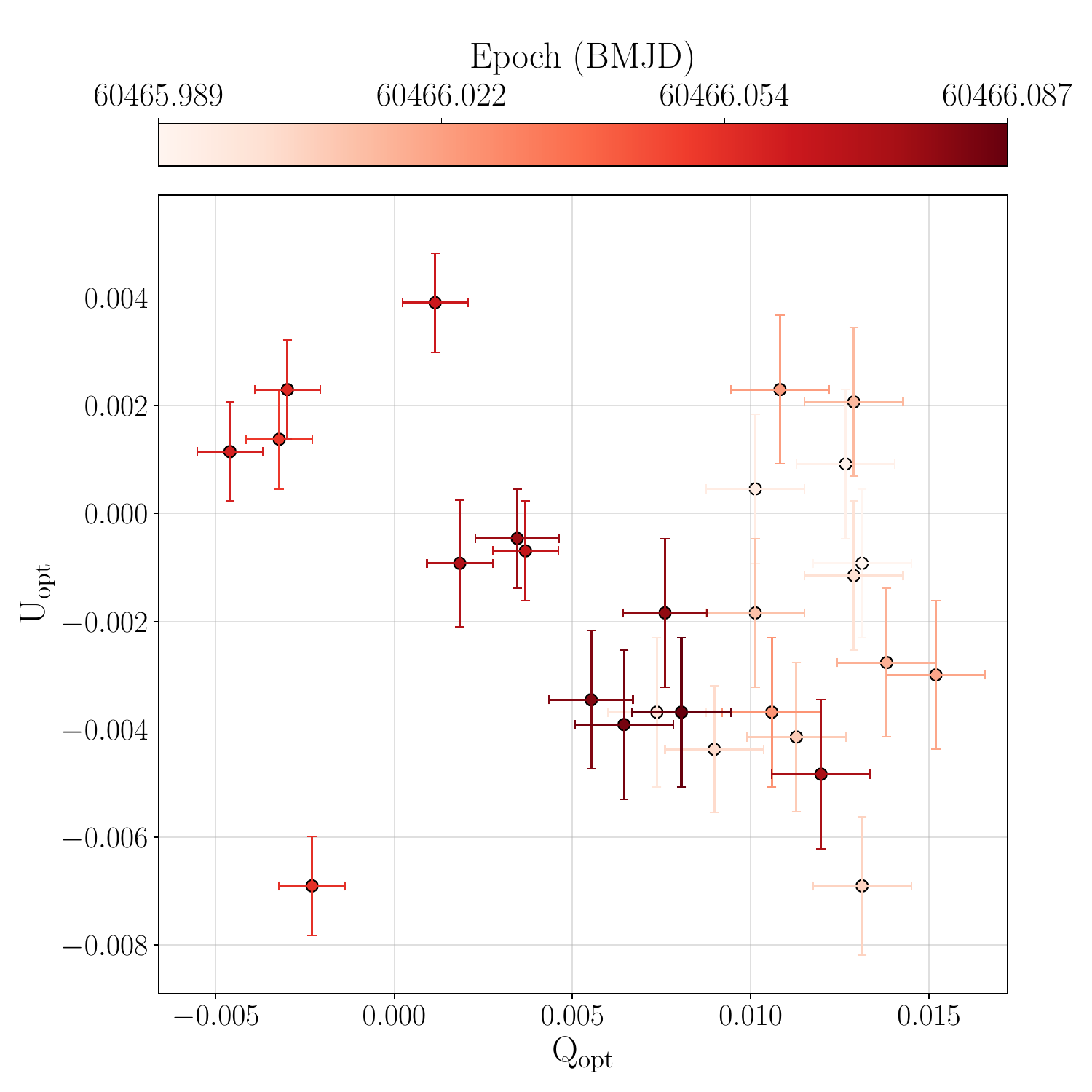}
\vspace{-0.5cm}
\caption{Stokes parameters $Q_{\rm opt}$ and $U_{\rm opt}$ for the optical emission. Data points are colour-coded according to the epoch of their measurements.}
\label{fig:Stokes_opt}
\end{center}
\end{figure}

The optical polarimetric analysis presented in this work follows that described by \cite{Baglio2020}. 
The VLT/FORS2 polarimeter works with a Wollaston prism that is inserted in the light path and splits the incoming light into two orthogonally polarised beams, and a rotating half-wave plate (HWP) captures images at four angles relative to the telescope axis ($\Phi_i = 22.5^{\circ}(i-1)$ for $i = 1, 2, 3, 4$). For our dataset, this procedure was repeated 28 times.
Images were reduced by subtracting an average bias frame and dividing by a normalised flat field; using {\tt daophot} \citep{Stetson1987}, aperture photometry was performed with a 6-pixel aperture. 
Once aperture photometry was performed on all stars in the field using {\tt daophot} \cite{Stetson1987}, we extracted the normalised Stokes parameters $Q_{opt}$ and $U_{opt}$ for linear polarisation as:

\begin{equation}\label{Q_U_eq}
Q= \frac{F(\Phi_1)-F(\Phi_3)}{2}; \,\,\, U=\frac{F(\Phi_2)-F(\Phi_4)}{2} ,
\end{equation}where

\begin{equation}
F(\Phi_i)=\frac{f^o(\Phi_i)-f^e(\Phi_i)}{f^o(\Phi_i)+f^e(\Phi_i)},
\end{equation}with $f^o$ and $f^e$ being the ordinary and extraordinary beams fluxes, respectively (see Sect.\,\ref{sec:vlt}).
The calculated Stokes parameters are shown in Fig. \ref{fig:Stokes_opt}, and are not corrected for instrumental polarisation. This correction can be achieved by observing a non-polarised standard star, as detailed in the FORS2 instrument documentation\footnote{\url{https://www.eso.org/sci/facilities/paranal/instruments/fors/tools/FORS1_Std.html}}. For FORS2, regular monitoring of unpolarised standard stars ensures accurate measurement of the instrumental polarisation. Over the past decade, these observations have demonstrated that the instrumental polarisation remains consistently low, within the range of 0--0.3\% across all bands\footnote{\url{http://www.eso.org/observing/dfo/quality/FORS2/reports/FULL/trend_report_PMOS_inst_pol_FULL.html}}.

We can then calculate the parameter $S(\Phi)$, which is the component of the normalised Stokes vector that describes the LP along the direction $\Phi$, for each HWP angle ($\Phi_i = 22.5^{\circ}(i-1)$ for $i = 1, 2, 3, 4$; Sect. \ref{sec:vlt})  as 
\begin{equation}\label{eq_S}
S(\Phi)=\left( \frac{f^{o}(\Phi)/f^e(\Phi)}{f^o_u(\Phi)/f^e_u(\Phi)}-1\right)\,\left( \frac{f^{o}(\Phi)/f^e(\Phi)}{f^o_u(\Phi)/f^e_u(\Phi)}+1\right)^{-1},
\end{equation}
where $f^{e}(\Phi)$ and $f^{o}(\Phi)$ are the extraordinary and ordinary fluxes at angle $\Phi$, and $f^e_u(\Phi)$ and $f^o_u(\Phi)$ are the corresponding fluxes for non-polarised standard stars (see \citealt{Baglio2024}).

$S(\Phi)$ relates to the polarisation level $P_{\rm opt}$ and angle $\theta_{\rm opt}$ as $S(\Phi) = P_{\rm opt}\, \cos 2(\theta_{\rm opt}-\Phi).$
$P_{\rm opt}$ and $\theta_{\rm opt}$ are derived using a two-step process (see \citealt{Baglio2020, Baglio2023}): initial estimates via Gaussian likelihood maximisation, and refinement of the values using Markov Chain Monte Carlo. The final values are taken as the median of the marginal posterior distributions, with uncertainties from the 16th - 84th percentiles.
This method corrects for instrumental and interstellar polarisation, since $S(\Phi)$ is normalised using non-polarised stars in the field (in this case, we selected 3 isolated stars in the field to enhance statistics). Given the low line-of-sight absorption and proximity of J1023, interstellar polarisation is likely well-subtracted with this method. A correction of $1.6\deg\pm 0.7\deg$ was applied on the polarisation angle using the polarised standard star Vela 1-95 (see \citealt{Baglio2024}).


\begin{figure}
\begin{center}
\includegraphics[width=0.48\textwidth]{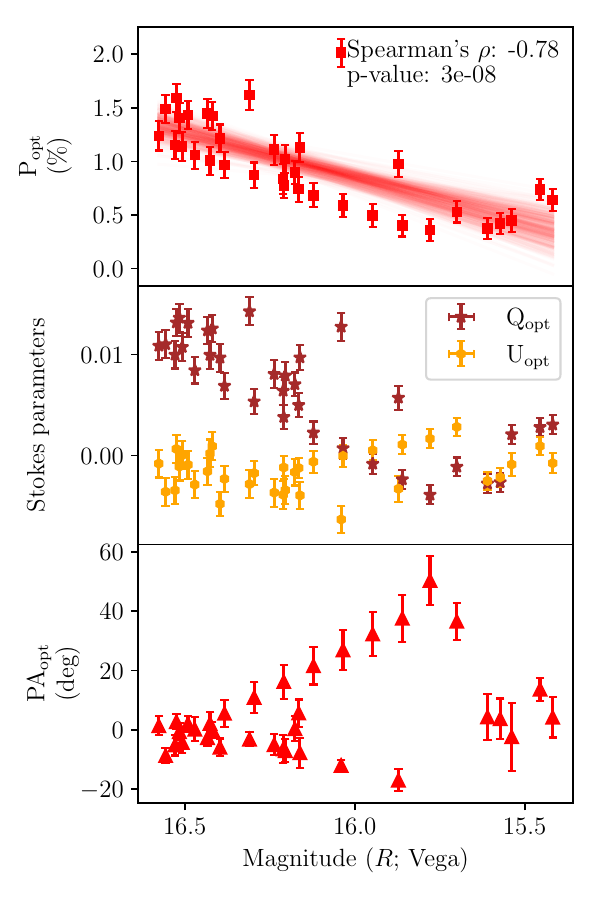}
\caption{Polarisation degree (top), Stokes parameters (middle), and polarisation angle (bottom) as a function of the relative intensity for the optical emission from J1023. The top panel shows the results of the Spearman rank correlation test and of a linear regression algorithm to the data. The shape of the anti-correlation is consistent with $\Delta$P$_{\rm opt}$ / $\Delta$mag = -1.0$\pm$0.2, with an intrinsic random scatter around the best regression model of $\sigma_0=0.07^{+0.04}_{-0.02}$ (see text for details).} 
\label{fig:flux_pol}
\end{center}
\end{figure}

Figure\,\ref{fig:flux_pol} shows the trend of $P_{\rm opt}$, $PA_{\rm opt}$, $Q$ and $U$ as a function of the $R$-band magnitude of J1023. A Spearman correlation test between $P_{\rm opt}$ and magnitude yielded a correlation coefficient of $\rho=-0.78$ and p-value of $3\times10^{-8}$ (see top panel of Fig.\,\ref{fig:flux_pol}). Using a Bayesian MCMC method with the \texttt{LINMIX}$_-$\texttt{ERR} model \citep{kelly07}, we derived a median slope of $\beta=-1.0\pm0.2$ and an intrinsic scatter of $\sigma_0=0.07^{+0.04}_{-0.02}$ from 10000 posterior samples. The middle panel of Fig.\,\ref{fig:flux_pol} shows that while $U_{\rm opt}$ remains roughly constant (around 0) during the flare, $Q_{\rm opt}$ is higher when the flux is lower. These results confirm that an unpolarised component becomes dominant during the flare.

\section{VLA observations}
\label{sec:Appendix_vla}

As in \cite{Baglio2024}, the VLA observations used the 3-bit sampler, dividing the 4096\,MHz bandwidth into 32 spectral windows with 64 2-MHz channels each. 3C286 served as the flux calibrator, J1024$-$0052 as the gain calibrator, and J1407$+$2827 as the leakage calibrator. J1023 was observed over 21 $\sim$9-min scans, totalling $\sim$3.15\,hr on-source.

We first processed the data using the \texttt{casa} VLA pipeline \citep[v6.5;][]{Casa2022}, removing and residual radio frequency interference (RFI) through a combination of manual editing and auto-flagging routings \texttt{rflag} and \texttt{tfcrop}. Currently, the VLA pipeline only calibrates the parallel-hand visibilities (i.e. Stokes $I$ and $V$). To measure the complete polarisation state, we performed additional calibration of the cross-hand visibilities (i.e. Stokes $Q$ and $U$) following the standard procedures\footnote{Tutorial: \url{https://casaguides.nrao.edu/index.php/CASA_Guides:polarization_Calibration_based_on_CASA_pipeline_standard_reduction:_The_radio_galaxy_3C75-CASA6.5.4}}. Stokes $IQUV$ images were created using \texttt{wsclean} \citep[v3.4;][]{wsclean}, applying a CLEAN-based deconvolution \citep{Hogbom1974}. Phase-only self-calibration was performed using \texttt{quartical} \citep{quartical}. The field of J1023 includes a bright radio galaxy (J102358.2$+$003826) 3$^{\prime\prime}$ away and other, fainter, point sources. We subtracted the visibilities of these background sources, allowing for narrow field-of-view images with minimal aliasing.

Given J1023's complex evolution, we created a suite of images to probe the range of phenomena; these included 20-second and 1-minute interval images to capture short-timescale variability and integrated images (i.e. by combining independent scans) for better signal-to-noise before, during, and after the flare. For each imaging interval, \texttt{wsclean} produced 16 channelised images and a frequency-averaged Multi-Frequency-Synthesis image. To measure the frequency-dependant time lags during the radio flaring, we also created a Multi-Frequency-Synthesis image for the lower (4--6\,GHz) and upper (6--8\,GHz) sub-bands. The Stokes $I$ flux density was measured using \texttt{imfit}, and for strong detections ($>20\sigma$), an intra-band spectral index was calculated. We also created linear polarisation images $(P_{\rm radio} = \sqrt{Q_{\rm radio}^2 +U_{\rm radio}^2})$. No polarised emission was detected, so we calculated $3\sigma$ upper limits following \cite{Vaillancourt2006}. 

\begin{figure}
    \centering
    \includegraphics[width=\columnwidth]{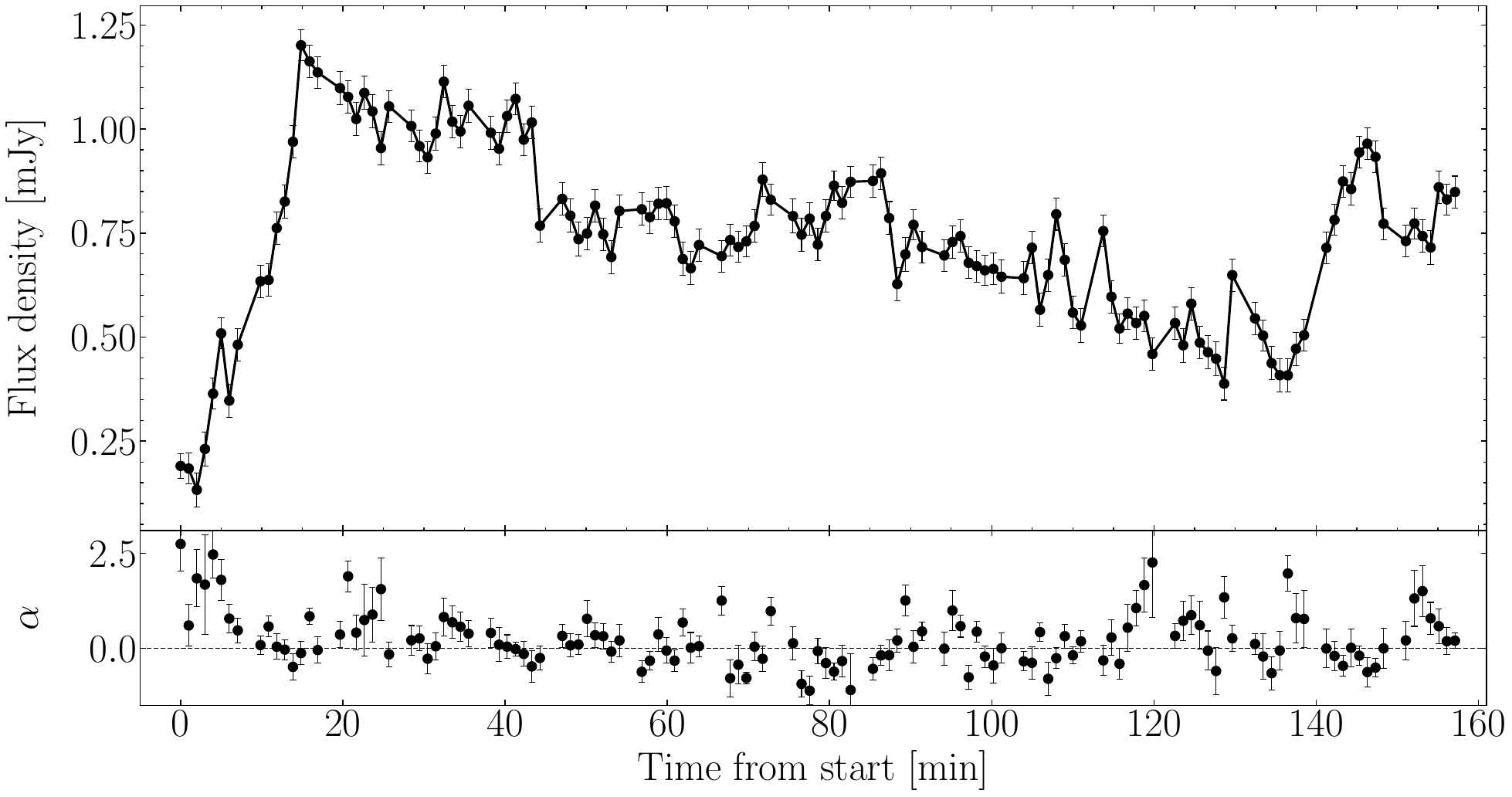}
    \vspace{-0.5cm}
    \caption{\textit{Top:} VLA 6\,GHz light-curve of J1023 from the September 2024 observation, with 1-minute time binning. The short-time scale variability from the flaring activity is evident. \textit{Bottom:} in-band spectral index with 1-minute time bins. The radio spectrum shows transitions between optically thick and thin phases on timescales of minutes.}
    \label{fig:VLA_SEP}
\end{figure}

We obtained an additional VLA observation of J1023 to check for evidence of propagation from any ejecta that may have been launched during the June flaring. The second VLA observation (DDT 24B-469) was conducted on September 15, 2024, between 14:51:18 and 17:29:27 UTC ($\sim$2\,hr on-source) with the same setup (C band, B configuration) and the same processing and reduction as for the June observation. However, the September observations did not include a leakage calibrator, so we exclusively analysed the Stokes $I$ properties.
The total exposure time on J1023 was 128 minutes, and the rms noise of the full observation was $\sim$2.5 $\mu$Jy. We obtained a clear detection of J1023 at $\sim$700 $\mu$Jy while it was displaying additional flaring activity, as shown in Fig.\,\ref{fig:VLA_SEP}. The flare shape and duration could be consistent with increased mass-loading of the compact jet or the launch of a discrete ejection, though current data cannot fully distinguish between these scenarios.

\section{Jet radiative luminosity}
\label{sec:radiative_lum}

As discussed in Sect.\,\ref{sec:Origin of the flaring behaviour}, we estimate the jet power to be $\simeq(5 - 8) \times 10^{32}$\,\lum, assuming that each flare from the first VLA observation results from a single ejection. Alternatively, if the radio flaring activity is due to increased mass loading in the compact jet, we can tentatively estimate the jet radiative luminosity by assuming a typical compact jet spectrum \citep{blandford79, russell20}. To perform this estimation, we integrate a flat spectrum (consistent with the average flare spectral index) between $10^9$\,Hz and the break frequency $\nu_{\rm break} = 2.5 \times 10^{13}$\,Hz (as estimated by \citealt{Baglio2023}), anchoring it at the VLA flux measured at 6\,GHz. Above the break frequency, we assume a typical optically thin spectrum with a spectral index of $\alpha = -0.6$ extending up to $10^{18}$\,Hz. We note that our results are only weakly dependent on this assumption. 

For the peak flux density of $\sim$1.5\,mJy during the flare, we obtain a total jet radiative luminosity of $\simeq 1.5 \times 10^{34}$\,\lum. We caution that this estimation is subject to significant uncertainty, as there are currently no measurements of the break frequency during the flaring mode. Notably, the jet luminosity in this mode is comparable to the source's accretion power in the high mode, $GM\dot{M}/R_{\rm LC} \approx 10^{32}-10^{34}$\,\lum, estimated considering an accretion rate of $\dot{M} \simeq 10^{13}-5 \times 10^{14}$\,g\,s$^{-1}$ and a light cylinder radius of $R_{\rm LC}$ = 80\,km \citep{linares22}.
For the radio flux before the flare ($\sim$$200$\,$\mu$Jy), we estimate a total jet radiative luminosity of $\simeq 2 \times 10^{33}$\,\lum. This value is again compatible with the accretion power in the high mode, considering the large (orders of magnitude) uncertainty in the mass accretion rate.

While it would be valuable to disentangle the contributions of different emission components through a spectral energy distribution, this is not feasible for the flaring mode. Capturing entire radio and optical flares would require long averaging, which would wash out spectral information on the radio flare substructures and combine X-ray data from different emission modes. This would most likely result in an oversimplified or misleading interpretation, as previous studies indicate that distinct emission components are very likely to be active in different modes.

Expanding on our comparison to V404 Cygni, the accretion luminosity from J1023 is five orders of magnitude lower than V404 Cygni in 2015 which, at the time of flaring, was accreting near the Eddington luminosity \citep[${\sim}\,10^{39}$\,\lum for a $\sim{10}M_\odot$ black hole;][and references therein]{Motta2017AKH}. In contrast, the jet power in each flare (assuming jet ejecta) is only separated by a factor of $\lesssim3000$; while very tentative, these results suggest that tMSPs in the flaring state may be more efficient at powering jet ejections than XRBs in their comparably luminous states (i.e. `the very high state'; \citealt{Done2007}).


\begin{figure}
\begin{center}
\includegraphics[width=0.67\columnwidth]{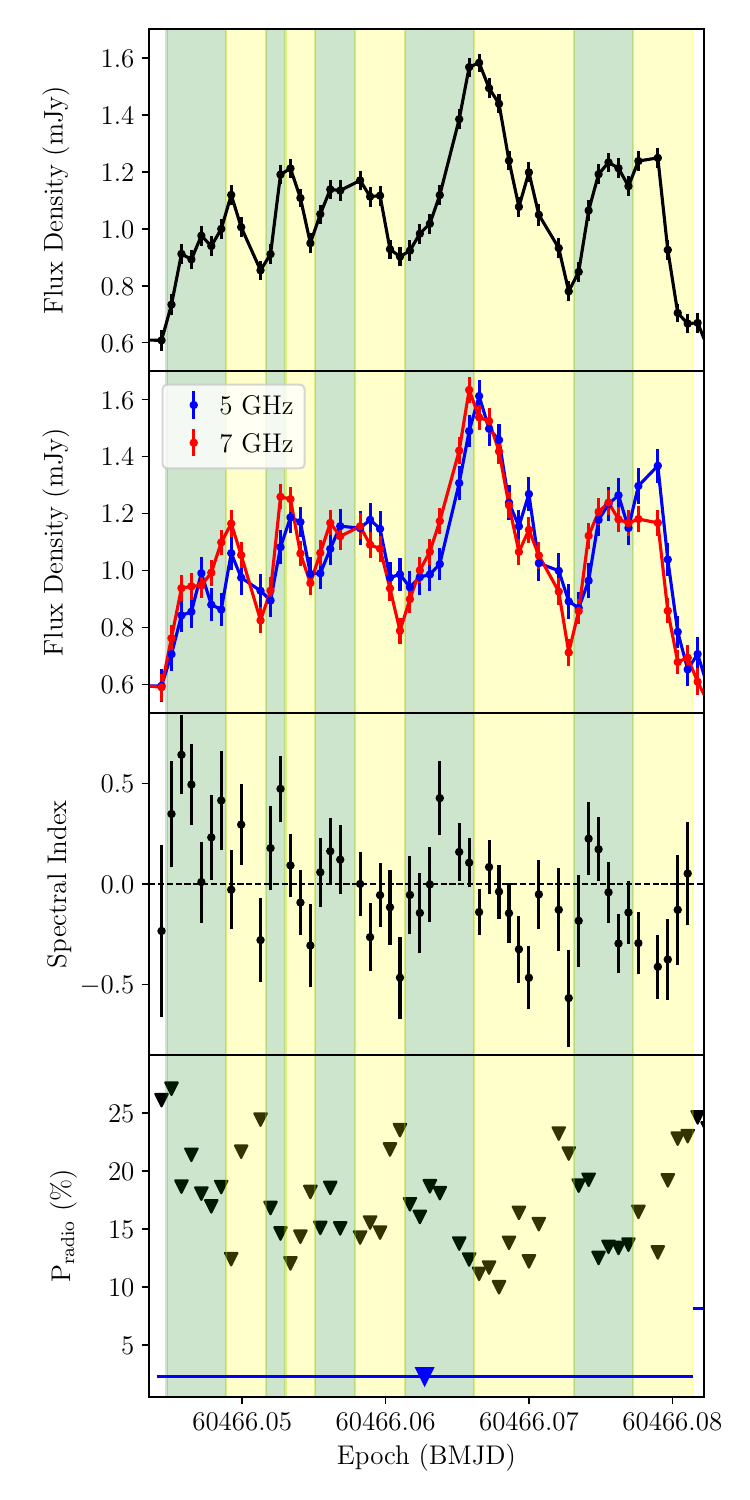}
\caption{\textit{Top:} VLA 6\,GHz light-curve of J1023 from the June 2024 observation during the flaring period, with 1-minute time binning. The radio emission shows multiple flares of similar duration, with green and yellow regions indicating flux rise and decay phases, respectively. \textit{Second panel:} The same light-curve split into two sub-bands centred on 5 and 7\,GHz. \textit{Third panel:} Evolution of the in-band spectral index $\alpha$ on 1-minute time scale. Here the radio flux density follows $S_{\nu} \propto \nu^{\alpha}$. Rising phases display flat-to-inverted spectra, while decay phases are generally optically thin. \textit{Bottom:} Time-resolved and integrated upper limit on the linear polarisation fraction of the radio emission.} 
\label{fig:spindex_flare}
\end{center}
\end{figure}
\begin{figure}
\begin{center}
\includegraphics[width=0.9\columnwidth]{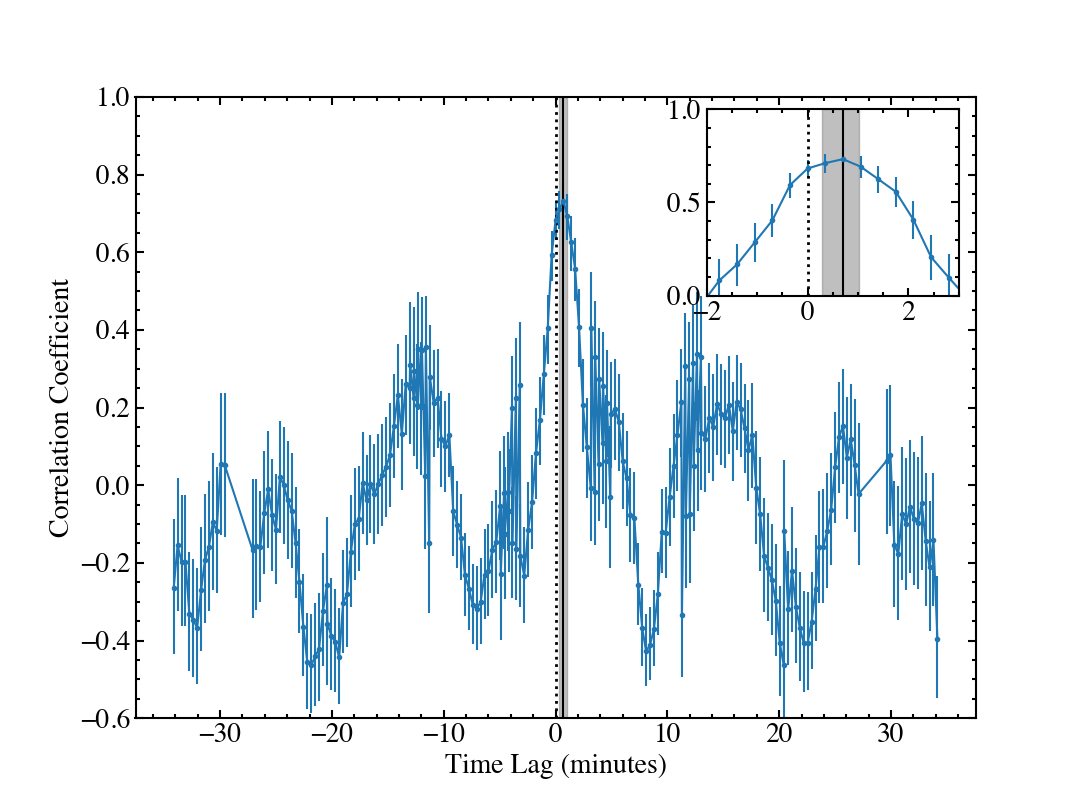}
\vspace{-0.2cm}
\caption{Cross-correlation function calculated with the \texttt{ZDCF} algorithm. We include the full range of time lags and a zoomed-in plot highlighting the peak lag. The solid vertical line and grey region show the best fit time lag and 1$\sigma$ uncertainties, respectively. The vertical dashed line shows the zero-lag position, offset from the best-fit lag at a 1.8$\sigma$ significance.} 
\label{fig:flare_delay}
\end{center}
\end{figure}

\end{appendix}

\end{document}